\documentclass[aps,prl,unsortedaddress,superscriptaddress,showpacs,twocolumn]{revtex4-1}
\usepackage{graphicx}
\usepackage{amssymb}
\usepackage{epstopdf}
\usepackage{subfigure}
\usepackage{hyperref}
\begin{document}
\title{Light sterile neutrino sensitivity at the nuSTORM facility}

\author{D.~Adey}
\affiliation{Fermi National Accelerator Laboratory, Box 500, Batavia, IL 60510-5011, USA}
\author{S.K.~Agarwalla}
\affiliation{Institute of Physics, Sachivalaya Marg, Sainik School Post, Bhubaneswar 751005, Orissa, India}
\author{C.M.~Ankenbrandt}\thanks{Also at Fermilab, P.O. Box 500, Batavia, IL 60510-5011, USA}
\affiliation{Muons Inc., 552 N. Batavia Avenue, Batavia, IL 60510, USA} 
\author{R.~Asfandiyarov}
\affiliation{University de Geneve, 24, Quai Ernest-Ansermet, 1211 Geneva 4, Switzerland}
\author{J.J.~Back}
\affiliation{Department of Physics, University of Warwick, Coventry, CV4 7AL, UK}
\author{G.~Barker}
\affiliation{Department of Physics, University of Warwick, Coventry, CV4 7AL, UK}
\author{E. Baussan}
\affiliation{IPHC, Universit\'e de Strasbourg, CNRS/IN2P3, F-67037 Strasbourg, France}
\author{R.~Bayes}\thanks{Corresponding author: Ryan.Bayes@glasgow.ac.uk}
\affiliation{School of Physics and Astronomy, Kelvin Building, University of Glasgow, Glasgow G12 8QQ, Scotland, UK}
\author{S.~Bhadra}
\affiliation{Department of Physics and Astronomy, York University, 4700 Keele Street, Toronto, Ontario, M3J 1P3, Canada}
\author{V.~Blackmore}
\affiliation{Oxford University, Subdepartment of Particle Physics, Oxford, UK}
% ...
\collaboration{The nuSTORM Collaboration}

\date{\today}

\begin{abstract}
  A facility that can deliver beams of electron and muon neutrinos
  from the decay of a stored muon beam has the potential to
  unambiguously resolve the issue of the evidence for light sterile
  neutrinos that arises in short-baseline neutrino oscillation
  experiments and from estimates of the effective number of neutrino
  flavors from fits to cosmological data.  In this paper, we show that
  the nuSTORM facility, with stored muons of 3.8 GeV/c $\pm$ 10\%, will be able
  to carry out a conclusive muon neutrino appearance search for
  sterile neutrinos and test the LSND and MiniBooNE experimental
  signals with 10$\sigma$ sensitivity, even assuming conservative
  estimates for the systematic uncertainties. This experiment would
  add greatly to our knowledge of the contribution of light sterile
  neutrinos to the number of effective neutrino flavors from the
  abundance of primordial helium production and from constraints on
  neutrino energy density from the cosmic microwave background.  The
  appearance search is complemented by a simultaneous muon neutrino
  disappearance analysis that will facilitate tests of various sterile
  neutrino models.
\end{abstract}

\maketitle

The issue of light sterile neutrinos is one of general interest to
particle physicists and cosmologists.  Intriguing evidence from
terrestrial neutrino sources exists for neutrino mixing between the
three active neutrinos and light sterile neutrino species. Short-baseline 
neutrino oscillations observed by the LSND
\cite{Aguilar:2001ty} and MiniBooNE \cite{Aguilar-Arevalo:2013pmq}
experiments, the collective evidence of the reactor neutrino anomaly
\cite{Huber:2011wv} and the gallium anomaly
\cite{Anselmann:1994ar,Hampel:1997fc,Abdurashitov:1996dp,Abdurashitov:1998ne,Abdurashitov:2005tb}
all point towards sterile neutrinos with masses at the electronvolt
level. While these results are tantalizing, they are not conclusive on
their own and there is tension with the disappearance searches, which
exclude the best-fit light neutrino \cite{Mahn:2011ea,Adamson:2011ku}.
Furthermore, estimates of the effective number of neutrino flavors
\cite{Asaka:2005an,Ade:2013zuv} from fits to cosmological data suggest
that this number is greater than three.  These results are based on
primordial helium production during big-bang nucleosynthesis and
constraints on neutrino energy density from the cosmic microwave
background. Assumptions based on the partial thermalization of the
primordial neutrino species \cite{Jacques:2013xr} and the inclusion of
uncertainties in the Hubble constant \cite{Gariazzo:2013gua} can be
used to accommodate all the available data.  Therefore, there is great
interest to resolve the issue of the existence of light sterile
neutrinos, with implications for particle physics and cosmology.  

New ideas have recently been proposed, based on carrying out
oscillation experiments from isotope decay-at-rest sources
\cite{Bungau:2012ys} and other accelerator, reactor and active source
neutrino experiments \cite{Abazajian:2012ys}.  In this letter we show
that the nuSTORM facility, providing neutrino beams from the decay of
muons in a storage ring, can unambiguously resolve the problem of the
existence of light sterile neutrinos by providing a source for all
short-baseline oscillation modes. This idea has evolved from previous
neutrino factory work carried out in the context of sterile neutrino
(2+2) and (3+1) models
\cite{Donini:1999jc,Donini:2001xy,PhysRevD.63.033002}.  We will show
in this letter that the currently proposed nuSTORM accelerator
facility is feasible, without the need for new technology, and that
the analysis presented is realistic, in terms of the detector
performance. The best sensitivity to sterile neutrinos can be achieved
with the $\nu_{e}\to\nu_{\mu}$ oscillation channel, conjugate to the
LSND measurement, but the simultaneous access to disappearance modes
can be used to test the consistency of the neutrino oscillation
hypothesis for the first time in a single experiment.


Muon decays in flight yield a neutrino beam with a precisely known
flavor content and energy spectrum. The primary decay mode,
$\mu^{+}\to e^{+}\nu_{e}\bar{\nu}_{\mu}$, is 98.6\% of all muon
decays. The remainder is made up of radiative decays, $\mu^{+}\to
e^{+}\nu_{e}\bar{\nu}_{\mu}\gamma$ (B.R.$\approx$1.4\%), and
$\mu^{+}\to e^{+}e^{-}e^{+}\nu_{e}\bar{\nu}_{\mu}$
(B.R. $=(3.4\pm0.4)\times10^{-5}$) \cite{Amsler:2008zz}. These
decays all have the same neutrino content (50\% $\nu_{e}$, 50\%
$\bar{\nu}_{\mu}$), so any difference in the neutrino flavor would
represent new physics. The energy spectrum of the muon decay positron
has been measured to be consistent with the standard model at
the level of a few parts in $10^{4}$ \cite{TWIST:2011aa}. 

\begin{figure}[bp]
  \begin{center}
    \includegraphics[width=\columnwidth]{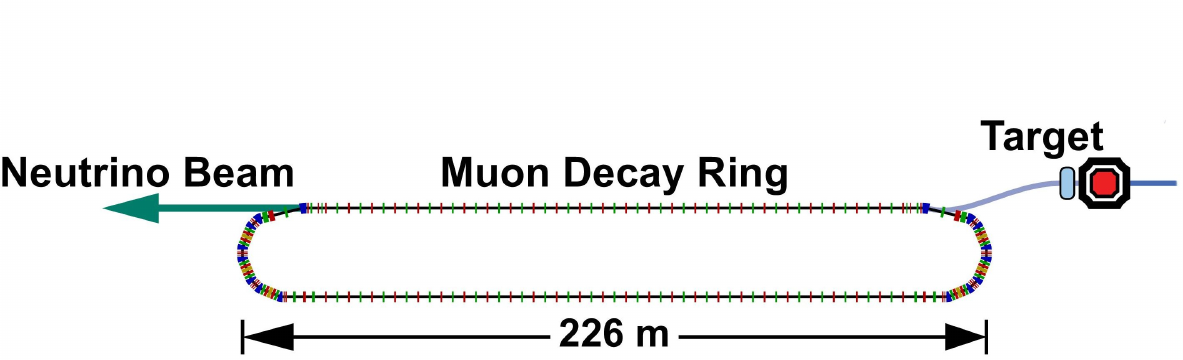}
    \caption{A schematic of the storage ring configuration. Pions are
      injected into a straight section and must decay into muons before the
      first bend or be ejected from the ring. Muons that decay in the
      injection straight during subsequent turns produce the neutrino beam.}
    \label{fig:schematic}
  \end{center}
\end{figure}

The nuSTORM facility has been designed to inject 5~GeV pions into a
muon storage ring \cite{Kyberd:2012iz}, with a beam lattice in a race
track configuration (Fig.~\ref{fig:schematic}). The effective straight
for neutrino production is 185~m long and includes pion injection and
extraction sections. The total circumference of the ring is 480~m. The
storage ring circulates muons with a central momentum of 3.8 GeV/c and
has a momentum acceptance of $\approx\pm$10\%. Pions that do not decay
prior to the first bend and muons produced from pion decay in the
forward direction are removed by an extraction section at the end of
the straight. Since the muons circulate many times between pion fills,
neutrinos from pion decay are separated from the sample of neutrinos
purely from muon decays through the use of a time cut that isolates
decays immediately after injection. Muons that decay in the bends or
in the opposing straight do not produce useful neutrinos. It is
expected that $\approx 2\times 10^{18}$ useful muon decays in the
production straight that points toward the far detector site can be
generated by nuSTORM from a total of $10^{21}$ protons on target (POT)
over a total of ten years~\cite{Kyberd:2012iz,Adey:2013afh}. The
neutrino beam has a dispersion of 29~mrad from the boost of the muon
decay and 4~mrad muon beam divergence in the production straight. The
uncertainty in the neutrino flux is expected to be less than 0.5\%,
due to the measurements to be carried out by beam monitoring devices
in the decay run and at a near detector.

\begin{figure}[hbp]
  \begin{center}
    \includegraphics[width=0.8\columnwidth]{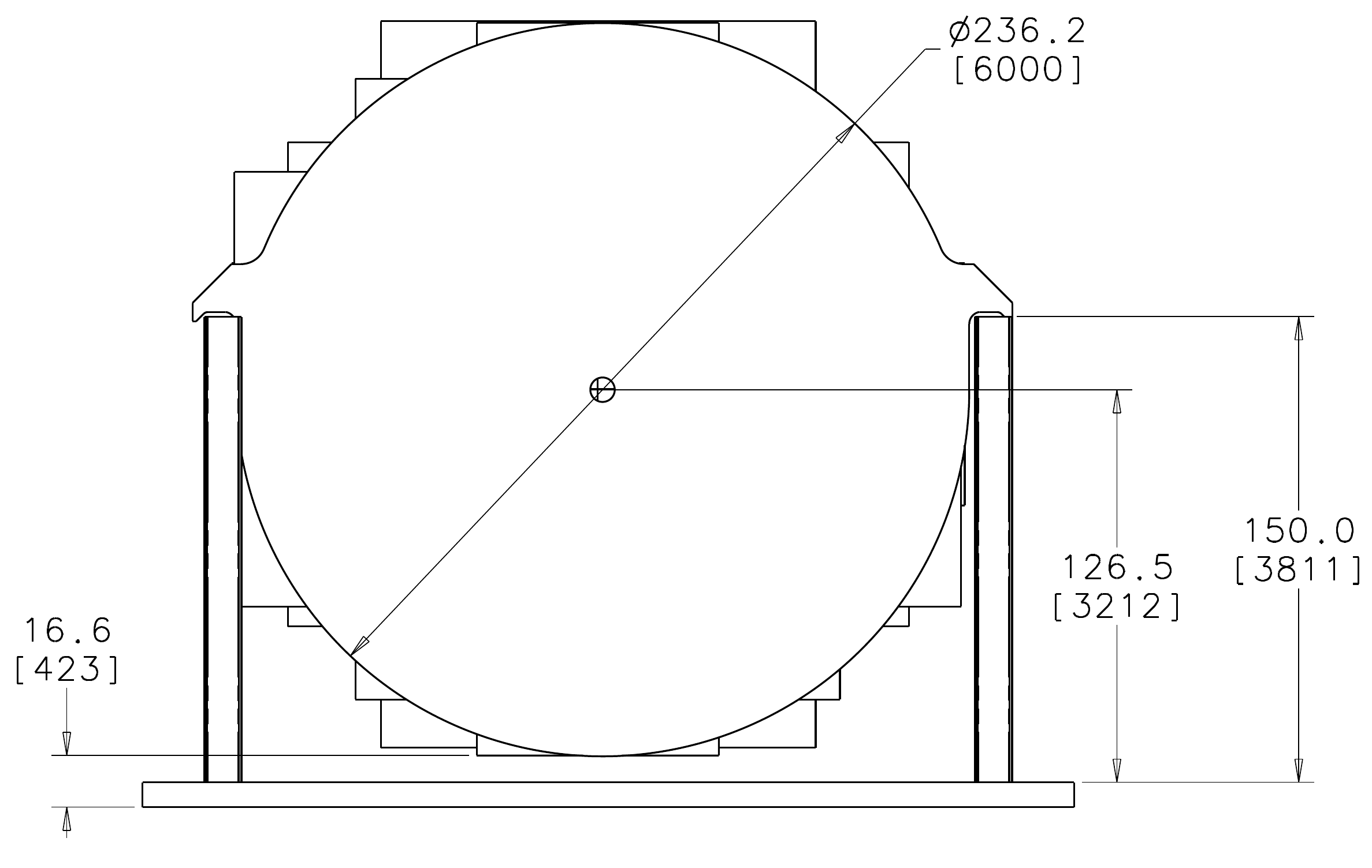}
    \caption{A cross section of the prospective iron-scintillator neutrino detector (6 m in diameter and 13 m in length).}
    \label{fig:detscheme}
  \end{center}
\end{figure}

The well defined neutrino beams available at the nuSTORM
facility grant unparalleled opportunities for neutrino physics. Rates
of accessible neutrino oscillation channels for stored $\mu^+$,
assuming a simple (3+1) sterile neutrino model~\cite{Kopp:2013vaa}
consistent with the LSND anomaly, are shown in
Table~\ref{tab:channels} for $10^{21}$ POT. The
probability of observing a $\nu_{e}\to\nu_{\mu}$ transition is
given by
\begin{equation}
P_{e\mu} = \sin^{2}2\theta_{e\mu}\sin^2\left(\frac{\Delta m^2 L}{4 E}\right)
\end{equation}
where $\theta_{e\mu}$ is an effective mixing angle, and $\Delta m^{2}$ is
the effective mass difference, independent of the sterile neutrino
model. In the (3+1) model $\sin^{2}2\theta_{e\mu} \equiv 4|U_{\mu
  4}|^2|U_{e4}|^{2}$ where $U_{\epsilon n}$ is an element of the PMNS
mixing matrix. Alternatively, the probability of  observing a 
$\bar{\nu}_{\mu}$ disappearance transition is given by
\begin{equation}
P_{\mu\mu} = 1 - \sin^{2}2\theta_{\mu\mu}\sin^2\left(\frac{\Delta m^2 L}{4 E}\right)
\end{equation}
where $\sin^{2}2\theta_{\mu\mu} \equiv 4|U_{\mu 4}|^{2}\left(1-|U_{\mu
  4}|^{2}\right)$ in a (3+1) model.

A $\nu_{\mu}$ appearance experiment is conducted by observing
$\mu^{-}$ in the detector and a $\bar{\nu}_{\mu}$ disappearance
experiment relies on identification of $\mu^{+}$ in the
detector. Therefore, the sensitivity to oscillations depends on the
ability of the detector to distinguish the charge of the leptons
produced in the neutrino charged current (CC) interactions. With the
rates shown in Table~\ref{tab:channels}, a background acceptance of
$10^{-4}$ is required for an appearance measurement. Direct
measurement of the cross-sections of both electron and muon neutrinos
can be measured at a near detector site 50~m from the end of the decay
straight. The number of $\nu_e$ and $\bar{\nu}_\mu$ CC events (per
100~ton fiducial mass at the near detector) is $4.0\times 10^6$ and
$2.1\times 10^6$, respectively, for a $10^{21}$ POT exposure.  It is
also possible to select $\mu^{-}$ in the storage ring. This will yield
a lower rate in the detection of appearance oscillations, and hence a
reduced sensitivity, due to the difference in the cross-section
between neutrinos and anti-neutrinos ($1.8\times 10^6$ $\bar{\nu}_{e}$
and $4.6\times
10^6$ $\nu_\mu$ CC events would be observed in the
near detector in this case).

\begin{table}
  \caption{Expected rates for neutrino oscillation channels observed at a
    1.3~kt detector, 2~km away from a muon storage ring with an
    exposure of 10$^{21}$ POT.}  
  \begin{tabular}{rccc}
    \hline
    Channel& Oscillation & $N_{osc.}$ & $N_{null}$  \\
    \hline
    $\nu_{\mu}$ Appearance & $\nu_{e}\to\nu_{\mu}$ CC& 332 & 0  \\
    $\bar{\nu}_{\mu}$ Disappearance & $\bar{\nu}_{\mu}\to\bar{\nu}_{\mu}$ CC &
    122322 & 128433 \\
    $\nu_{e}$ Disappearance & $\nu_{e}\to\nu_{e}$ CC & 216657 & 230766 \\
    NC Disappearance & $\bar{\nu}_{\mu}\to\bar{\nu}_{\mu}$ NC & 47679 & 50073 \\
    NC Disappearance &   $\nu_{e}\to\nu_{e}$ NC & 73941 & 78805 \\
 
    \hline
  \end{tabular}
    \label{tab:channels}
\end{table}

A 1.3~kt magnetized iron-scintillator calorimeter has been selected as
the detector for short-baseline oscillation physics at nuSTORM, 
as it has excellent charge selection and
detection characteristics for muons. This 6~m diameter detector is to
be constructed of modules of 1.5~cm thick steel plates, and two layers
of scintillator bars to yield 3D space points at each measurement
plane. The overall length of this detector is 13~m. Each scintillator
bar has a cross-section of 2.0$\times$0.75~cm$^{2}$ and will be read
out using silicon photo-multipliers. For a schematic of this detector,
see Fig.~\ref{fig:detscheme}. The magnetic field will be generated by
a 240~kA-turns current carried by 8 turns of a super-conducting
transmission line. This provides a toroidal
magnetic field between 1.9 and 2.6 Tesla within the steel.


A detailed simulation of the iron-scintillator far detector was
developed from the neutrino factory far detector simulation
\cite{Bayes:2012ex}. This simulation uses the GENIE
\cite{Andreopoulos:2009rq} package to simulate neutrino interactions
in iron and scintillator, and GEANT4 \cite{Apostolakis:2007zz} to
simulate the interactions of the products with the detector
material. A simple digitization is used to group ionization sites to
particular paired scintillator bars and replicate the effects of
resolution and attenuation within the scintillator bars. Tracks are
reconstructed from the events through repeated application of a Kalman
filter \cite{CerveraVillanueva:2004kt} to determine the momentum and
charge of tracks. Multiple tracks are fit from each event and the
longest track is defined as the muon. Other tracks, if present, are
assumed to be the result of pion production and other particle shower
processes. A track is reconstructed from pion or shower events in 1\%
of cases, necessitating further analysis to remove such events.

\begin{table}[b]
\caption{Variables used in the definition
  of the classifier for the multi-variate analysis of events in the detector
  simulation. }
	\begin{tabular}{rp{5cm}}
	\hline\hline
	Variable & Description \\
	\hline
	Track Quality & $\sigma_{q/p}/(q/p)$, the normalized error in the track curvature.\\
	Hits in Trajectory & The number of sci. planes in track. \\
	Curvature Ratio & $(q_{init}/p_{range}) \times
        (p_{fit}/q_{fit})$: ratio of the initial estimate and
        Kalman fit momentum. \\
	Mean Energy Deposition & $\sum^{N}_{i=0}\Delta E_{i}/N$ for
        planes in track. \\
	Variation in Energy & $\sum^{N/2}_{i=0}\Delta E_{i} / \sum^{N}_{j=N/2}\Delta E_{j}$, where the energy deposited per hit $\Delta E_{i}~<~\Delta E_{i+1}$.\\
	\hline\hline 
	\end{tabular}
\label{tab:var}
\end{table}%

\begin{figure}
  \subfigure[Appearance efficiencies]{
    \includegraphics[width=\columnwidth]{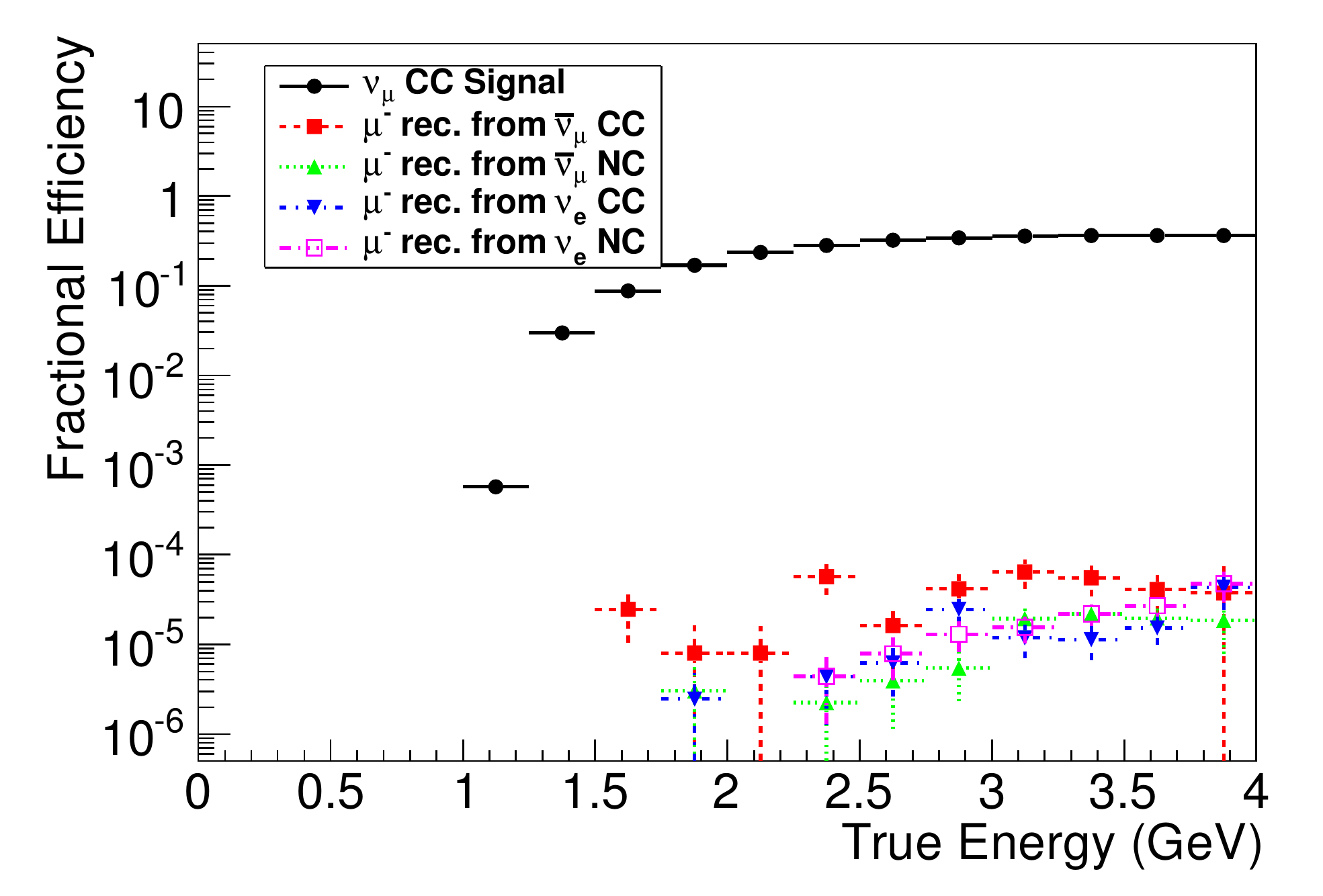}
    \label{fig:effapp}
  }
  \subfigure[Disappearance efficiencies]{
    \includegraphics[width=\columnwidth]{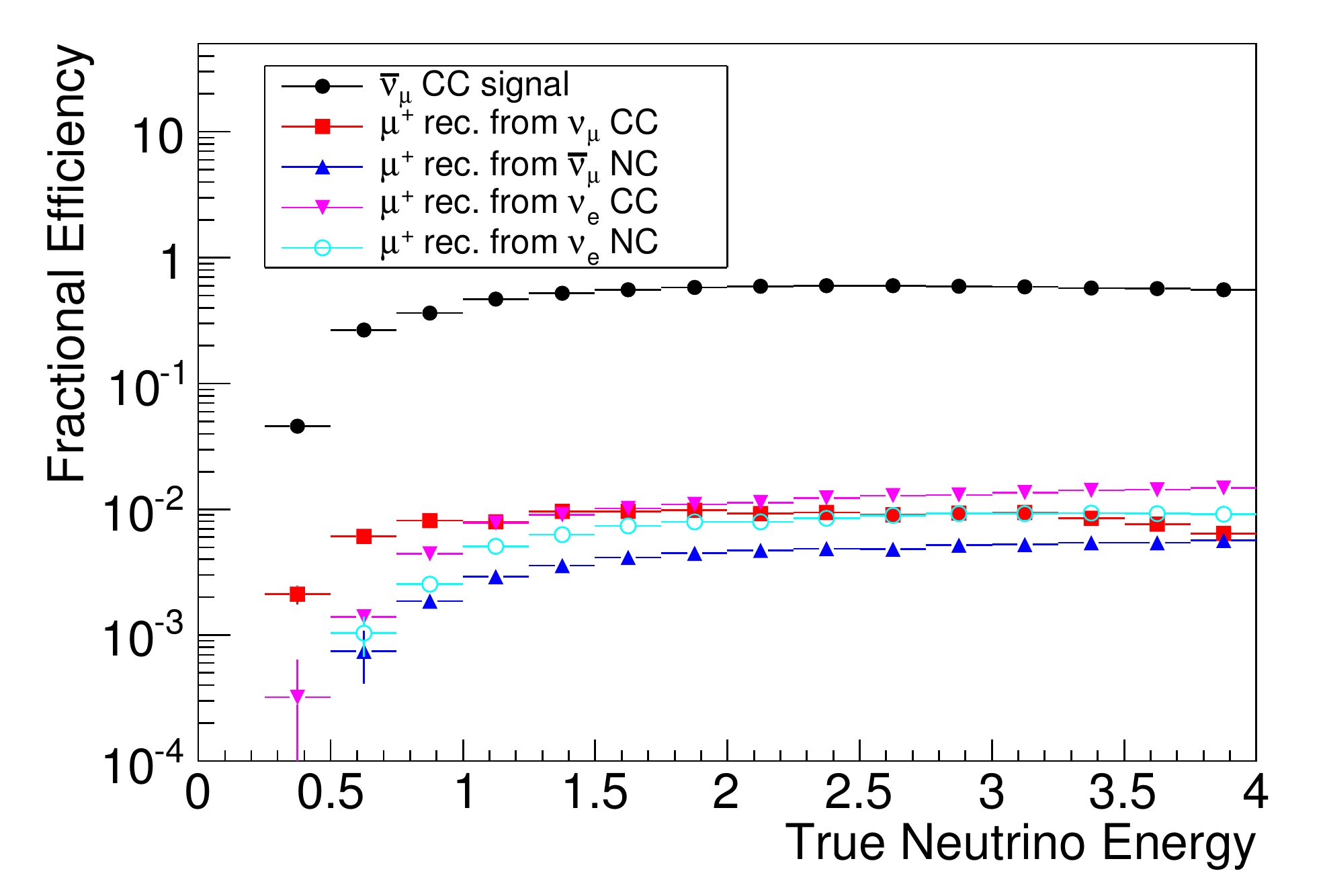}
    \label{fig:effdis}
  }
  \caption{Efficiencies of signals and backgrounds for $\nu_{\mu}$
    appearance and $\bar{\nu}_{\mu}$ disappearance for an
    iron-scintillator neutrino detector optimized for the region of
    interest for nuSTORM. The appearance analysis
    used a BDT algorithm to define the above curves, while the
    disappearance analysis used a neural network (MLPBNN) as described
    in the text.}
  \label{fig:eff}
\end{figure}

A multi-variate analysis of reconstructed events is used to
distinguish signal events from background with a high degree of
purity. A series of cuts to perform a pre-selection of events were
applied first.  These are based on finding one or more tracks in the
event; on successfully fitting the longest track; imposing a maximum
momentum $p_{\mu}<4$~GeV/c for the longest track; applying a fiducial
cut in which the longest track must start before the last 1~m of the
detector; 60\% of the hits assigned must be associated to the longest
track; the relative error on the ratio $q/p$, where $q$ and $p$ are
the fitted charge and momentum of the track, must satisfy
$\sigma_{q/p}/(q/p) < 10.0$; and the ratio of the initial curvature
over the fitted curvature satisfies $(q_{init}/p_{init}) \times (p/q)
> 0$.  An approach was used based on the boosted decision tree (BDT)
algorithm provided by the TMVA \cite{Hocker:2007ht} subset of the
ROOT~\cite{Brun:1997pa} analysis package in which five track variables
(shown in Table~\ref{tab:var}) are used to discriminate between muons
from $\nu_{\mu}$~CC interactions and all other types of
interactions. The method reduces these five track variables to one
classifier variable that runs between 0 and 1, based on a training
process that differentiates between $\nu_{\mu}$~CC events, the
experimental signal, and $\bar{\nu}_{\mu}$~NC events, representing the
experimental backgrounds. The trained multivariate analysis (MVA) is
applied to simulations corresponding to the entries in
Fig.~\ref{fig:effapp} to determine the detector response to signal
($S$) and background ($B$) events. Given the expected number of
oscillated and unoscillated neutrinos at the far detector, an optimal
signal significance --- quantified as $S/\sqrt{S+B}$ --- is achieved
for an appearance experiment when the classifier is restricted to
values greater than 0.86. This yields an integrated signal efficiency
of 0.17 and a background efficiency of $4\times 10^{-5}$. This
background is predominantly due to charge mis-identification from
$\nu_{\mu}$ CC events, but also contains pion decay and punch-through
from NC events. A cuts-based analysis was also studied
\cite{Kyberd:2012iz,TunnellThesis}, based on the number of hits in a
trajectory and the track quality, but yielded a decreased physics
sensitivity, with a signal efficiency of 0.16 and a background
efficiency of $5\times10^{-5}$ at a higher energy threshold.

For a disappearance analysis, a different optimization is required
since background rejection is a lesser concern. An optimization using
a $\chi^{2}$-statistic between neutrino spectra, given the (3+1)
sterile neutrino hypothesis and the standard neutrino hypothesis,
concludes that a neural network (MLPBNN) algorithm
\cite{Hocker:2007ht} that retains classifier values greater than 0.94
outperforms the BDT algorithm. The efficiency curves for the optimized
analysis are shown in Fig.~\ref{fig:effdis}.


The detector response for each class of event shown in
Fig.~\ref{fig:eff} is extracted from the detector simulation as a
``migration'' matrix of the probability of a neutrino generated in the
$i^{\rm th}$ energy bin being reconstructed in the $j^{\rm th}$ energy
bin. The migration matrices are input into a simulation of the
oscillation experiment using the GLoBES software package
\cite{Huber:2004ka} with modifications to simulate non-standard
interactions \cite{Kopp:2013vaa} and accelerator effects, such as the
integration of muon decays from positions throughout the decay
straight \cite{TunnellThesis,Tunnell:2012nu}. The GLoBES simulations
assume an experiment with a 1.3~kt far detector at a distance of 2~km
from the end of the storage ring, with 1.6$\times 10^{18}$ useful muon
decays. The total appearance signal is 73 events, with a combined
background of 6 events, assuming $\Delta m_{14}^{2}=0.89$~eV$^2$ and
$\theta_{14}=0.15$~rad.

\begin{table}[tp]
  \centering
  \caption{Systematic uncertainties expected for a short-baseline
    muon neutrino appearance experiment based at nuSTORM.}
  \begin{tabular}{rcc}
  \hline
    Uncertainty & 
    \multicolumn{2}{c}{Expected Contribution} \\
    & Signal & Background \\
    \hline
    Flux & 0.5\% & 0.5\% \\
    Cross section &  0.5\% & 5\%  \\
    Hadronic Model &  0 & 8\% \\
    Electromagnetic Model &  0.5\% & 0 \\
    Magnetic Field & $0.5$\% & $0.5$\%\\
    Variation in Steel Thickness & 0.2\%  & 0.2\% \\
    \hline
    Total & 1\% & 10\%\\
    \hline
  \end{tabular}
  \label{tab:sys}
\end{table}

\begin{figure}[htbp]
\begin{center}
\includegraphics[width=\columnwidth]{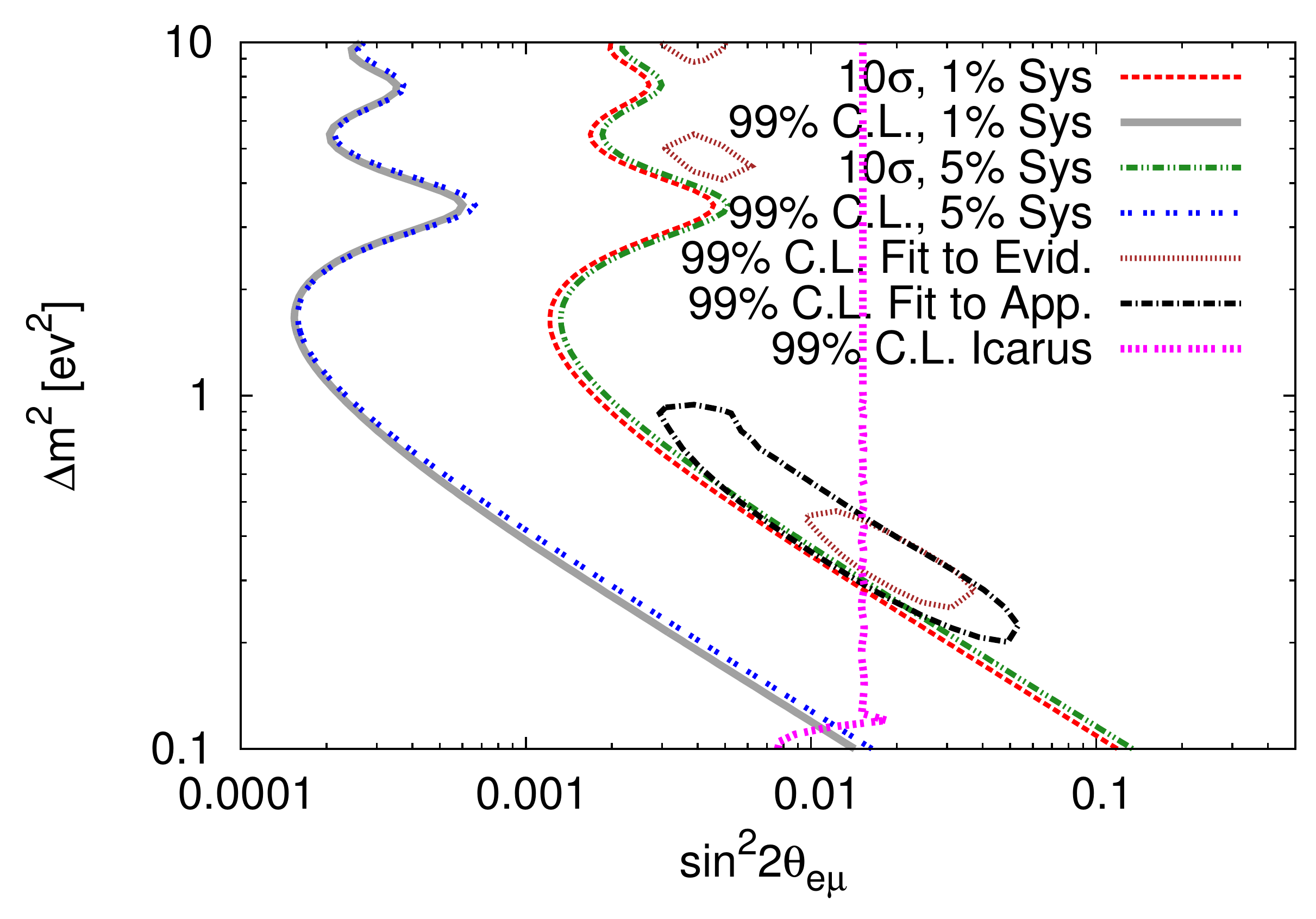}
\end{center}
\caption{Sensitivity of nuSTORM to the $\nu_{e}\to\nu_{\mu}$
  appearance oscillation due to the presence of sterile neutrinos
  assuming a (3+1) model with anticipated and inflated systematics,
  compared to 99\% confidence contours from global fits to the
  evidence for sterile neutrinos and to all available appearance
  experiments generated by Kopp {\it{ et. al.}}  \cite{Kopp:2013vaa}
  (filled contours) and limits set by ICARUS
  \cite{Antonello:2013gut}.}
\label{fig:appsig}
\end{figure}

\begin{figure}
  \includegraphics[width=\columnwidth]{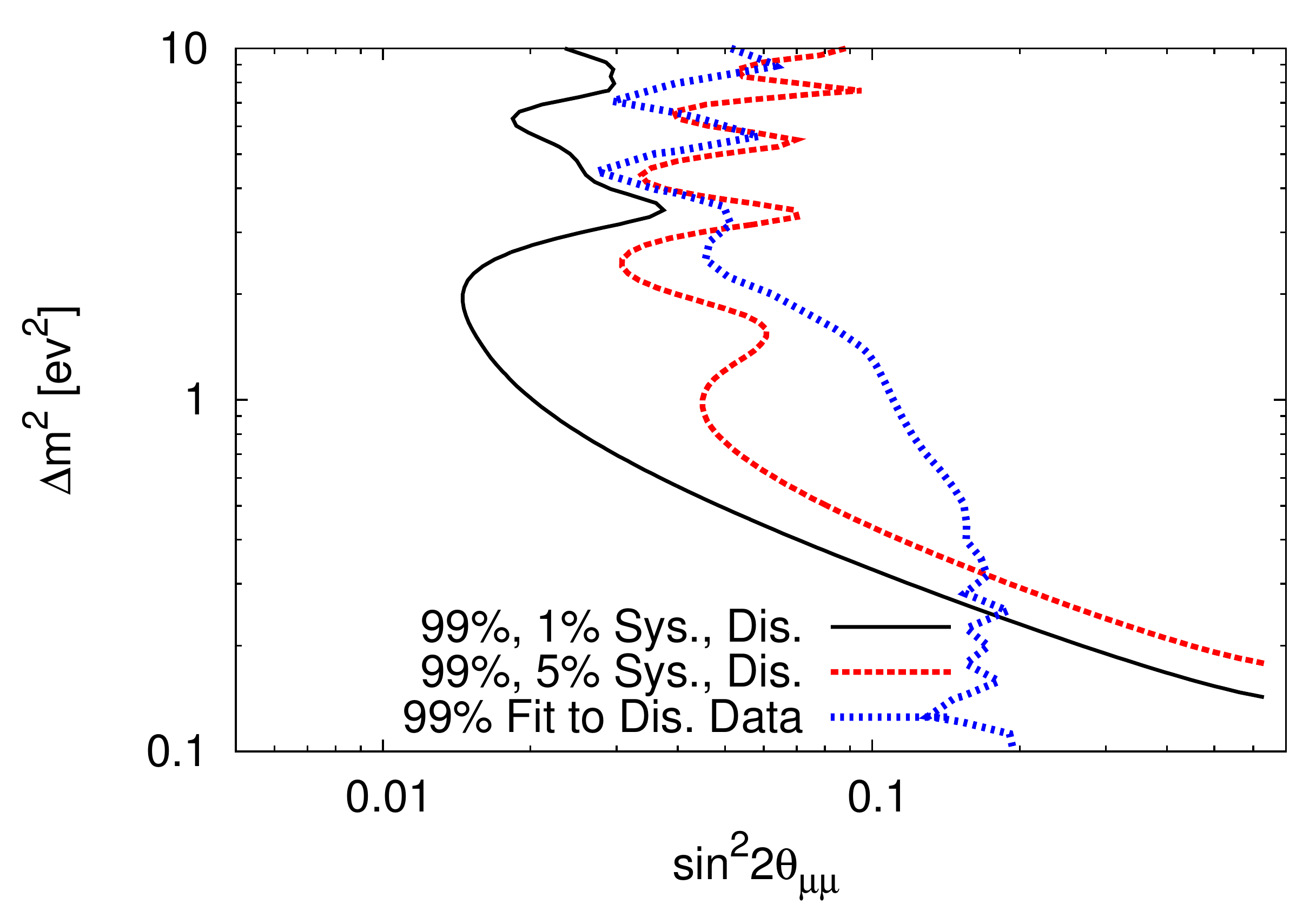}
  \caption{Sensitivity of nuSTORM to $\nu_{\mu}$ disappearance oscillations
    assuming a (3+1) neutrino model. Contours are generated from a
    $\chi^{2}$ statistic assuming both the anticipated and inflated
    systematic uncertainties compared to the exclusion contour produced from the fit of
    $\sin^{2}2\theta_{\mu\mu}$ and $\Delta m^{2}$ to the existing
    disappearance data (blue dots) \cite{Kopp:2013vaa}.}
  \label{fig:dissig}
\end{figure}

The sensitivity of a $\nu_{\mu}$ appearance experiment to the presence
of sterile neutrinos in a (3+1) model as a function of $\Delta
m_{14}^{2}$ and $\sin^{2}2\theta_{e\mu}$ is shown in
Fig.~\ref{fig:appsig} assuming the anticipated systematic
uncertainties (Table~\ref{tab:sys}) and systematic uncertainties
inflated to 5\% (signal) and 50\% (background), using a boosted
decision tree analysis. This is compared to the 99\% confidence
contours from fits generated by Kopp {\it{ et. al.}}
\cite{Kopp:2013vaa} to the combination of LSND, MiniBooNE, and the
reactor and gallium disappearance experiments (``Fit to Evid.''), and to all
available appearance data (``Fit to App.'') and to the recent 99\%
C.L. contour from the long-baseline ICARUS experiment
\cite{Antonello:2013gut}, neglecting matter effects.

Neutrino cross-section uncertainties can be reduced by direct
measurements conducted with the beams produced by nuSTORM in both the
$\nu_\mu$ and $\nu_e$ channels.  For the appearance experiment,
relative systematic uncertainties due to differences in cross-sections
of neutrino and anti-neutrino, and electron and muon neutrinos will
primarily affect the backgrounds, and therefore are strongly
suppressed. The uncertainty in the quasi-elastic scattering
cross-section relative to the total cross section will affect the
signal and the background equally. Such measurements will greatly
contribute to the physics in the neutrino generators used for
reference simulations. However, as the appearance search is a rate
limited measurement, energy calibration effects such as the known
GENIE model uncertainties \cite{Coloma:2013tba} should not affect the
results described here.

The sum of these systematic uncertainties will yield a total 1\%
uncertainty to the total normalization of the signal and a 10\%
uncertainty to the background. In the absence of any such
measurements, an upper limit can be taken from existing experiments,
such as MINOS \cite{Adamson:2010uj}. The convolution of the flux
multiplied by the cross-section, based on current MINOS data, was 
used to determine the uncertainties to be 4\% for signal and
40\% for background. For an upper bound to the sensitivity of the
described experiment, inflated uncertainties of 5\% and 50\% are
considered. The appearance experiment is still sensitive to the
presence of a sterile neutrino consistent with the existing evidence
at the 10$\sigma$ level, as shown in Fig.~\ref{fig:appsig}. Cosmic ray
backgrounds were also considered through the application of the CRY
software package \cite{Hagman:2012}. With the application of
self-vetoing cuts on the fiducial volume to a skin depth of 30~cm, the
cosmic ray background is reduced to less than 1 event per year.

A simultaneous and statistically independent $\bar{\nu}_{\mu}$
disappearance measurement will be conducted with the same experimental
setup. Sensitivity contours as a function of $\Delta m_{14}^{2}$ and
$\sin^{2}2\theta_{\mu\mu}$ are shown in Fig.~\ref{fig:dissig}. A near
detector is essential to extrapolate the expected neutrino flux at the
far detector~\cite{Laing:2008zzb,Choubey:2011zzq}. It is assumed that
the systematic uncertainties used in the appearance measurement are
the same as those for the disappearance measurement. The
$\bar{\nu}_{\mu}$ disappearance measurement is far more sensitive to
systematic uncertainties due to the increase of the signal and
background acceptance. The exclusion contours set by the nuSTORM
disappearance measurement alone shows improvement in the 99\%
C.L. bounds over the current global fits as in
Fig.~\ref{fig:dissig}. The true sensitivity is expected to fall
between the pessimistic and optimistic cases, because the inclusion of
the flux extrapolation from the 200 Tonne near detector is expected to
introduce similar systematic uncertainties while it constrains the
spectral uncertainty. The simulation of the near detector required to
test this assertion is in progress. An optimization of a $\nu_e$
disappearance experiment at a similar muon storage ring facility with
idealized detector systems was carried out, demonstrating the near-far
extrapolation \cite{Winter:2012sk}, but the realistic assessment of
this channel is still in progress.

The presence of light sterile neutrinos, consistent with the
short-baseline neutrino anomalies and from estimates of the effective
number of neutrino flavors that arise from fits to cosmological data,
would provide evidence for physics beyond the Standard Model and would
have far-reaching consequences in neutrino physics and cosmological
models of large structure formation.  In this letter, we have
demonstrated that the nuSTORM facility can deliver high purity beams
of neutrinos to carry out a $\nu_e$ to $\nu_{\mu}$ neutrino
oscillation appearance measurement, using an iron-scintillator
calorimeter detector at a distance of 2~km, with a signal significance
of better than 10$\sigma$. The simultaneous use of the
$\bar{\nu}_{\mu}$ disappearance channel grants nuSTORM added potential
to resolve the current tension between appearance and disappearance
measurements and potential to resolve differences between sterile
neutrino models.  The experimental sensitivity of the appearance channel
is largely robust to systematic effects. Therefore, this experiment
would be able to provide the definitive test for light sterile
neutrinos and resolve a long-standing problem.

\bibliography{nuSTORM}

\end{document}